\documentclass[aps,prb,twocolumn,letterpaper,superscriptaddress,showpacs]{revtex4}
\usepackage{CJK}
\usepackage{keyval}%
\usepackage{graphicx}
\usepackage{dcolumn}
\usepackage{bm}
\usepackage{color}
\usepackage{amsmath}
\usepackage{soul}

\setkeys{Gin}{width=0.8\columnwidth,clip=true}

\newcommand{\ignore}[1]{}

\begin{document}
\begin{CJK*}{UTF8}{bsmi}
\title{Avoiding critical-point phonon instabilities in two-dimensional materials: The origin of the stripe formation in epitaxial silicene}
\author{Chi-Cheng Lee (%
李啟正
)}
\affiliation{School of Materials Science, Japan Advanced Institute of Science and Technology (JAIST),
1-1 Asahidai, Nomi, Ishikawa 923-1292, Japan}%
\author{Antoine Fleurence}
\affiliation{School of Materials Science, Japan Advanced Institute of Science and Technology (JAIST),
1-1 Asahidai, Nomi, Ishikawa 923-1292, Japan}%
\author{Rainer Friedlein
}
\affiliation{School of Materials Science, Japan Advanced Institute of Science and Technology (JAIST),
1-1 Asahidai, Nomi, Ishikawa 923-1292, Japan}%
\author{Yukiko Yamada-Takamura
}
\affiliation{School of Materials Science, Japan Advanced Institute of Science and Technology (JAIST),
1-1 Asahidai, Nomi, Ishikawa 923-1292, Japan}%
\author{Taisuke Ozaki
}
\affiliation{School of Materials Science, Japan Advanced Institute of Science and Technology (JAIST),
1-1 Asahidai, Nomi, Ishikawa 923-1292, Japan}%
\affiliation{Institute for Solid State Physics, The University of Tokyo, Kashiwa 277-8581, Japan}%

\date{\today}

\begin{abstract}
The origin of the large-scale stripe pattern of epitaxial silicene on the ZrB$_2$(0001) surface observed by scanning tunneling microscope experiments is revealed by first-principles calculations. Without stripes, the ($\sqrt{3}\times\sqrt{3}$)-reconstructed, one-atom-thick Si layer is found to exhibit a ``zero-frequency'' phonon instability at the $M$ point. In order to avoid a divergent response, the relevant phonon mode triggers the spontaneous formation of a new phase with a particular stripe pattern offering a way to lower both the atomic surface density and the total energy of silicene on the particular substrate. 
The observed mechanism is a way for the system to handle epitaxial strain and may therefore be more common in two-dimensional epitaxial materials exhibiting a small lattice mismatch with the substrate.   
\end{abstract}

\pacs{63.22.-m, 64.70.Nd, 68.35.Ja, 68.43.Fg}

\maketitle
\end{CJK*}

Ultimately thin, two-dimensional (2D) materials are considered top candidates for future electronic technologies since they offer high charge carrier mobilities\cite{Novoselov,Geim} and among many others interesting optical and vibrational properties\cite{Louie,Maultzsch,Chou,Flores,Yevick,FengLiu}. Their reduced dimensionality is sometimes intimately connected to lattice instabilities\cite{Flores,Yevick,FengLiu,Schrieffer,Ross,Shirane,Ghosez}, to the nesting of Fermi surfaces \cite{Cr} and to exciting electronic and spintronic phenomena\cite{Little,Cr,Kane,MacDonald}. When placing 2D materials on surfaces, underlying mechanisms might then be complex, especially considering that complicated surface reconstructions\cite{Smeu}, coexisting multiple structural phases of the layers\cite{Qikun} and/or domain(-like) structures\cite{Chen} are often observed. 

In this context, recently, a promising new 2D material called silicene\cite{Takeda,Ciraci} has attracted great attention. Due to the mixed $sp^{2}/sp^{3}$ bonding configurations and varying interactions with so far metallic substrates, experimentally realized epitaxial silicene phases have been demonstrated to possess a variety of both geometrical and electronic structures\cite{Kawai,Vogt,Fleurence,Feng,Chen,Meng}. Different to silicene phases on Ag(111) surfaces\cite{Kawai,Vogt,Feng,Chen,Meng}, epitaxial silicene forms \textit{spontaneously} through surface segregation on the zirconium diboride (0001) surface grown on Si(111) substrates\cite{Fleurence}. For this system, the ($\sqrt{3}\times\sqrt{3}$) reconstruction of silicene represents itself as a ($2\times2$) reconstruction of the ZrB$_2$(0001) surface and is connected to a buckling on the atomic length scale\cite{Fleurence}. 

\begin{figure}[tbp]
\includegraphics[width=1.00\columnwidth,clip=true,angle=0]{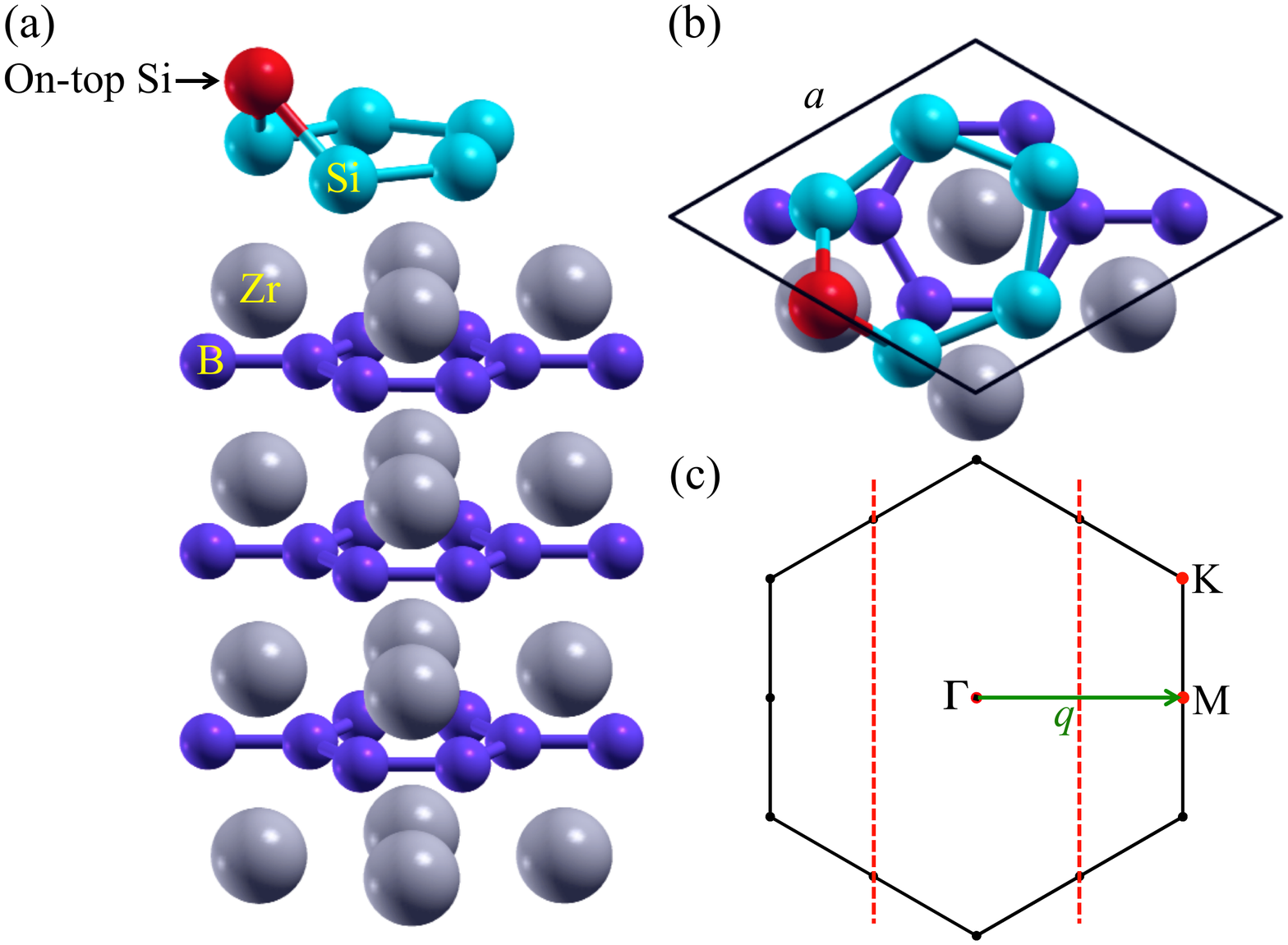}
\caption{\label{fig:fig1}
(a) Bird's eye view and (b) top view of the planar-like structure model of silicene on the ZrB$_2$(0001) surface. (c) First BZ of ($\sqrt{3}\times\sqrt{3}$)-reconstructed silicene. The length of the wave vector $q$ connecting the $\Gamma$ and $M$ points is $\frac{2\pi}{\sqrt{3}a}$ where $a$ is the calculated in-plane lattice constant of bulk ZrB$_2$.
}
\end{figure}

A structure model called the ``planar-like phase'' possessing only a single protruding Si atom per hexagon, shown in Figs.~\ref{fig:fig1} (a) and (b), has been reported to be the ground state within density functional theory\cite{Chicheng}. The band structure has been found to be in striking agreement with angle-resolved photoelectron (ARPES) spectra\cite{Friedlein,Chicheng2} providing solid support for the presence of this particular structural form of silicene. Consequently, when discussing epitaxial silicene on ZrB$_2$(0001), in the following, we refer to this ground-state structure model. Note that, as our yet unpublished calculations show, the other phase described as the ``buckled-like'' structure model\cite{Chicheng} possesses imaginary phonon frequencies and is thus unstable at least if the entropy term in the Helmholtz free energy\cite{Eriksson} is not considered. 

Note that while the evidence for the presence of the ``planar-like phase'' is strong, there has still been a major unresolved issue, namely the appearance of large-scale periodic stripe patterns observed by scanning tunneling microscopy (STM)\cite{Fleurence}. Note in this context that the planar-like form of silicene has also been suggested to be one of the available structures on the Ag(111) surface\cite{Chen} where no stripe pattern has been observed. Quite obviously, resolving the mechanism underlying the pattern exclusive to silicene on ZrB$_2$(0001) thin film surfaces prepared on Si wafers might provide new insight into the interplay between epitaxial conditions, strain and structural and electronic properties of two-dimensional adlayers in general.    

In this Letter, we report the results of a first-principles study of silicene on ZrB$_2$(0001). The driving force behind the mechanism that triggers the appearance of the stripe pattern is found to be associated with a phonon instability characterized by a zero-frequency mode reaching a critical point of instability. Dictated by the particular substrate, this mode cannot gain energy by following the corresponding eigendisplacement since the energy surface is essentially flat. This then immediately bears the question of how the lowest-energy state of the Si layer on the ZrB$_2$(0001) surface can be reached. 
We are going to demonstrate that the formation of a stripe pattern characterized by a distinct periodicity and by the reduction of the surface atom density is the way to obtain the lowest total energy and to avoid the phonon instability.   

In our approach to the phonon calculations, the dynamical matrix has been constructed from real-space force constants calculated using the OpenMX code which is based on norm-conserving pseudopotentials generated with multi reference energies\cite{MBK} and optimized pseudoatomic basis functions\cite{Ozaki,openmx}. The $(4\times4)$ ZrB$_2$(0001) supercell and 0.02 \AA\ for each atomic displacement have been adopted. One Si-, four Zr-, and three B-layers have been chosen to form an isolated slab implemented by the effective screening medium method\cite{Ohwaki}. A cut-off energy of 220 Ry has been used for numerical integrations and for the solution of the Poisson equation. The in-plane lattice constant has been obtained from the theoretical value of bulk ZrB$_2$ within the generalized gradient approximation (GGA)\cite{Chicheng,Kohn,Perdew}. The atomic positions have been relaxed to be less than $6\times10^{-5}$ Hartree/Bohr. For the quality of $k$-point sampling, an $8\times8$ mesh for the ($2\times2$) ZrB$_2$(0001) unit cell has been chosen. For each of the Si and B atoms, two and two optimized radial functions and a single one have been allocated for the $s$-, $p$-, and $d$-orbitals ($s2p2d1$), respectively. For each Zr atom, $s3p2d2$ is adopted. For all basis functions, a cut-off radius of 7 Bohr has been chosen. The XCrySDen software has been used to generate the figures\cite{xcrysden}.

\begin{figure}[center]
\includegraphics[width=1.00\columnwidth,clip=true,angle=0]{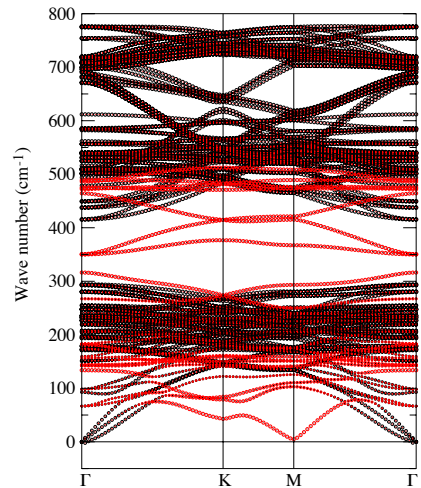}
\caption{\label{fig:fig2}
Phonon dispersion relation of silicene on ZrB$_2$(0001). The contributions of silicene and of the ZrB$_2$ substrate are presented by red and black circles, respectively. The diameters of circles represent the corresponding absolute squares of the atomic components in the eigenvectors.  
}
\end{figure}

The phonon dispersion relation of silicene on ZrB$_2$(0001) is presented in Fig.~\ref{fig:fig2}. No imaginary frequencies are identified. This suggests that silicene in its planar-like form is overall stable with zirconium diboride as the substrate. However, the softest vibrational mode reaches the ``zero'' frequency at the $M$ point. This high-symmetry point thus represents a \textit{singular point} of phonon instability that could cause a divergent response. Note that the presence of the critical-point phonon mode and therefore of the instability exists \textit{without} asserting additional external strain. Instead, it is related to the epitaxial strain imposed onto silicene by the particular zirconium diboride thin film surface which has a small lattice mismatch of about 5\% to the predicted value of hypothetical, free-standing silicene\cite{Takamura,Chicheng}. 
 
In order to find preferences for silicene structures away from the planar-like phase without stripes, the nature of the zero-frequency mode at the $M$ point should be understood. Before looking at this particular mode, it is worth mentioning that, except at the frequencies around 200 cm$^{-1}$, the phonon modes of silicene are almost decoupled from those of the ZrB$_2$ substrate. Since especially for the optical branches with high frequencies, those vibrations involve only Si atoms, it is suggested that the effect of the substrate to those phonons consists simply in providing a \textit{fixed} potential. Moreover and importantly, the softest, critical-point phonon mode that belongs to one of the acoustic branches also exhibits strong contribution from the Si atoms of silicene. It is then plausible that the silicene formation mechanism is governed by how the \textit{individual} Si atoms experience the interaction with the Zr-terminating surface of ZrB$_2$(0001). Specifically, the Si atoms located right on top of Zr atoms possess the least favorable in-plane positions as compared to Si atoms located in the hollow and bridge sites\cite{Chicheng}. 

\begin{figure}[tbp]
\includegraphics[width=1.00\columnwidth,clip=true,angle=0]{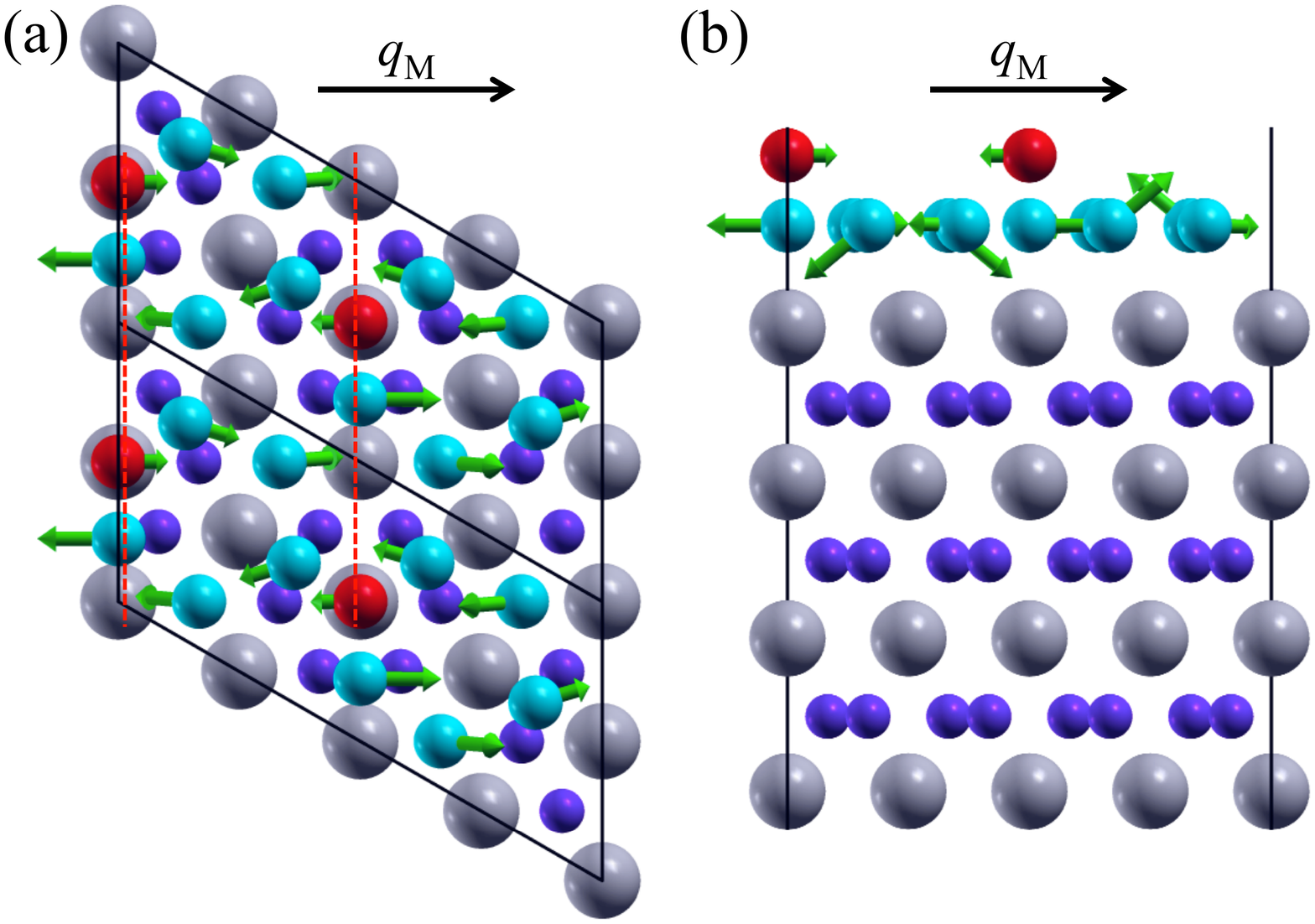}
\caption{\label{fig:fig3}
(a) Top view and (b) side view of the eigendisplacement of the ``zero''-frequency phonon mode at the $M$ point. The magnitude of the displacement is proportional to the length of the arrow. The deviations of on-top Si atoms is indicated by dashed lines. The direction of the wave vector $q_{M}$ is also indicated.  
}
\end{figure}

The eigendisplacement of the softest mode at the $M$ point is shown in Fig.~\ref{fig:fig3}. In this vibration, the on-top Si atoms move as guided by the dashed lines in Fig.~\ref{fig:fig3}(a). Since on-top positions are energetically unfavorable\cite{Chicheng}, Si atoms can gain energy when being displaced which allows for the potential energy to be nearly similar to that in the equilibrium positions, \textit{e.g.} for an almost \textit{flat} potential energy surface. As a result, this particular vibrational mode does require only little or no energy to be excited. Hence, the perfectly repeated planar-like structure cannot epitaxially be formed as a stable structure on ZrB$_2$(0001). 

A simple way to avoid instability would be to simultaneously elongate the in-plane lattice constants of both the diboride thin film substrate and the epitaxial silicene thereby releasing unfavorable in-plane stress\cite{Fleurence,Chicheng}. For the given ZrB$_2$ thin film substrate with its fixed lattice constants, however, silicene would be \textit{incommensurate} with the ($2\times2$) unit cell of the ZrB$_2$(0001) surface. Such a structure is energetically unfavorable since a simple expansion of the silicene lattice itself would lead to deviations of atomic positions away from the designated on-top, near-bridge and hollow Si sites. In order to maintain the planar-like structure and the commensurate relationship with the substrate to the highest degree possible, an obvious way is to form domains in which the commensurate relationship can be maintained around the domain centers and for which more flexible boundaries could allow for a modification of the crucial wave vector related to the $M$ point, $q_M$, such that the phonon instability might be avoided. How the system would ``engineer'' the Brillouin zone (BZ) of silicene is explained in the following. 

The key to unlock the size of the new BZ can be found by considering lines that connect on-top Si atoms, shown in Fig.~\ref{fig:fig3} (a). The wave length associated with $q_M$ can simply be modified by removing the on-top atoms in some of these lines thus creating a boundary with a lower surface density of Si atoms that separates stripes from each other. The associated enlargement of the repeat unit along the direction perpendicular to the axis of the stripes will shift the $M$ point of silicene. Therefore, the zone boundaries of silicene and of the ZrB$_2$(0001) surface become mismatched even in the repeated zones due to the slightly shorter $\Gamma$-$M$ wave vector of silicene. By doing so, the new $M$ point climbs out of the bottom of the dispersion valley away from the critical-frequency point. 

While the enlargement of the unit cell in one direction will apparently affect the wave lengths along the same direction, the symmetry-related $M$ points in the other directions should be considered as well. To avoid the phonon instability associated with all of the $M$ points simultaneously, the system may develop an \textit{effective smaller BZ} that is within the area described by dashed lines shown in Fig.~\ref{fig:fig1} (c) in which all of the original $M$ points are explicitly avoided. Then the width of the BZ corresponding to each stripe should be smaller than half of the original $\Gamma$-$M$ wave vector. Accordingly, the size of the unit cell related to the minimal stripe width should be larger than $2\sqrt{3}a$, four times the lattice constant of the ($2\times2$) ZrB$_2$(0001)-reconstructed surface projected on the $\Gamma$-$M$ direction.        

\begin{figure}[tbp]
\includegraphics[width=1.00\columnwidth,clip=true,angle=0]{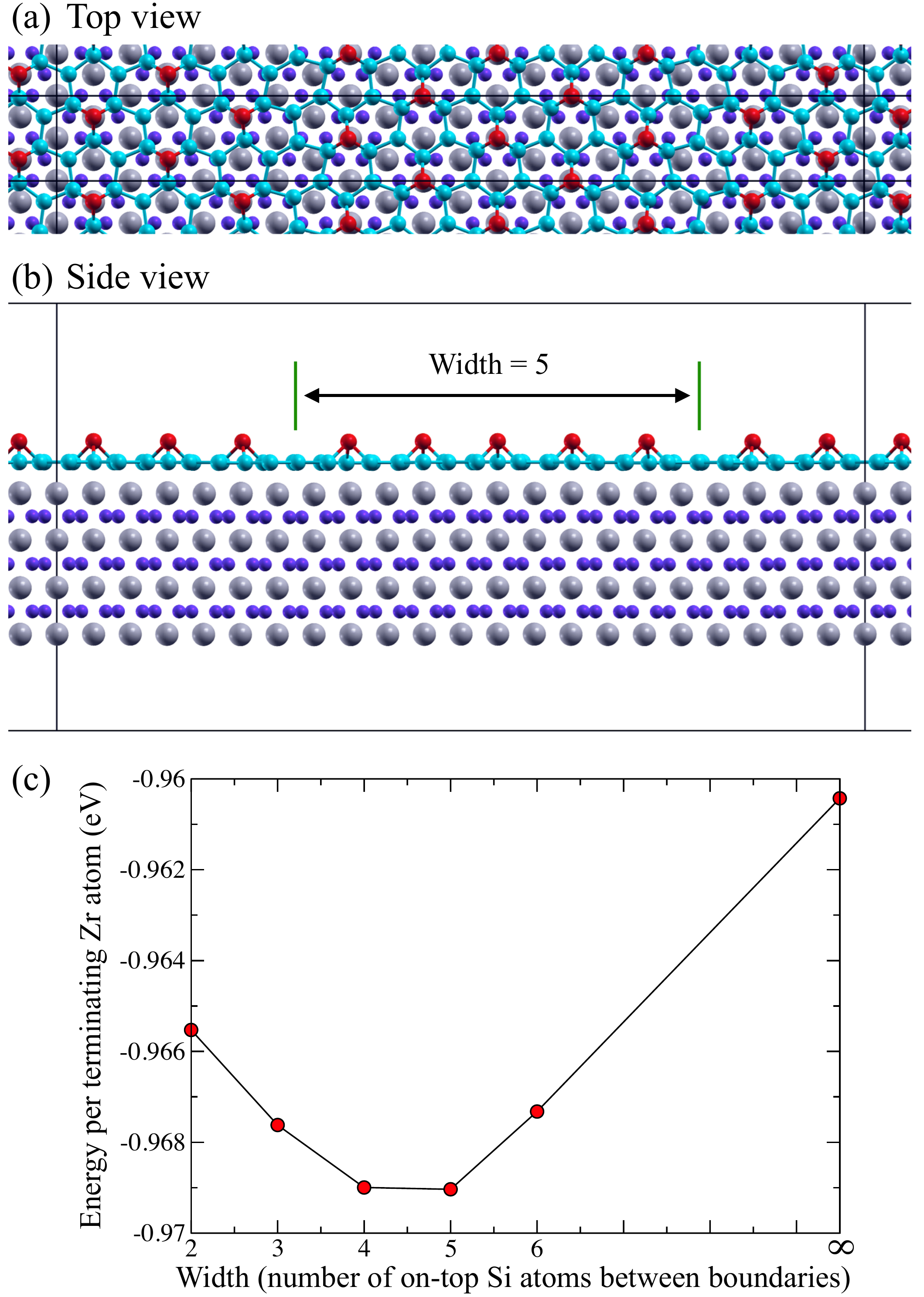}
\caption{\label{fig:fig4}
(a) Top view and (b) side view of fully relaxed atomic positions of the structure model for the stripe width of five on-top Si atoms. Bonds between Si atoms are shown to guide the deformed hexagons at the boundaries. (c) The formation enthalpy per terminating Zr atom versus the stripe width. 
}
\end{figure}

While so far, based on physics arguments, we derived conditions that determine the optimal stripe width, in the following, one set of possible large-scale structures will be explored. As shown in Figs.~\ref{fig:fig4} (a) and (b), boundaries are introduced by removing lines of on-top Si atoms. Two stripes separated by the respective boundaries possess different orientations of locally ($\sqrt{3}\times\sqrt{3}$)-reconstructed silicene. Note that, importantly, it is then still possible to connect the stripes by (distorted) hexagons such that the overall honeycomb-like silicene layer is maintained. Following the full relaxation of atomic positions, the corresponding structure calculated for the stripe width of five on-top Si atoms is shown in Figs.~\ref{fig:fig4} (a) and (b). Similar structures are found for other stripe widths. 

Any definition of the formation enthalpy of silicene on the ZrB$_2$(0001) thin film surface that includes structure models with varying stripe widths involves different surface densities of Si atoms. In order to reasonably estimate the energy gain per unit area, we therefore define the formation enthalpy as $E$(silicene $+$ ZrB$_2$ thin film) - $N$(silicene) $\times$ $E$(one Si atom) - $E$(ZrB$_2$ thin film) divided by $N$(terminating Zr layer). $E$ and $N$ denote the total energy and number of atoms, respectively. Here, $E$(one Si atom) refers to the total energy of Si atoms in the diamond structure since in the surface segregation process occurring during the formation of epitaxial silicene, the atoms needed to form silicene stem from the Si(111) substrate wafer underneath the diboride thin films. 

In Fig.~\ref{fig:fig4} (c) is plotted the formation enthalpy per terminating Zr atom as a function of the stripe width. 
The formation enthalpies are negative which explains why silicene is epitaxially formed on the ZrB$_2$(0001) surface.  
At least for the stripe widths of in between two and six on-top Si atoms,  silicene with stripes is more stable than silicene without stripes which corresponds to infinitely width. The most desirable width are stripes consisting of five on-top Si atoms across stripes although the energy gain is almost degenerate to the case with four on-top Si atoms. 
This confirms that (\textit{i}) the stripe pattern is needed, that (\textit{ii}) the  optimal width is not as short as possible, and (\textit{iii}) that the calculated results are highly consistent with the physical understanding discussed above. In addition, the sensitivity of the phonon frequency to the particular wave length also explains why stripe patterns do not show up for silicene phases prepared on other substrates\cite{Chen}. 

In conclusion, the study of the phonon instability of ($\sqrt{3}\times\sqrt{3}$)-reconstructed epitaxial silicene on the ZrB$_2$(0001) thin film surface uncovers the decisive factors that trigger the formation of the large-scale stripe pattern observed in STM images. 
In order to avoid the critical point of phonon instability, silicene spontaneously develops stripes by forming boundaries that miss on-top Si atoms. In particular, the surface is found to ``engineer'' the Brillouin zone by modifying the periodic boundary condition thus avoiding the critical wave vector and phonon frequency at the singular $M$ point. This is different to the case of charge density waves originating from the electronic energy gain at the Fermi energy upon nesting of the Fermi surface. Here, the large-scale lattice distortions in form of ``domain walls'' provide the energy gain in the order of phonon frequencies. Unlike for structural phase transitions where new structures are precisely coupled to the wave lengths of the associated soft modes, the demonstrated stripe width is chosen to avoid the $M$ point instability. By forming stripes, the ($\sqrt{3}\times\sqrt{3}$)-reconstructed planar-like silicene is largely preserved. Additionally, the observed mechanism of the critical-point-driven large-scale stripe formation within a single layer provides understanding of how epitaxial two-dimensional materials can efficiently and delicately handle epitaxial strain caused by a small lattice mismatch.

We are grateful for the use of the Cray XC30 machine at JAIST. This work is supported by the Strategic Programs for Innovative Research (SPIRE), MEXT, the Computational Materials Science Initiative (CMSI), and Materials Design through Computics: Complex Correlation and Non-Equilibrium Dynamics, A Grant in Aid for Scientific Research on Innovative Areas, MEXT, Japan. We also acknowledge financial support from the Funding Program for Next Generation World-Leading Researchers (GR046).

\bibliography{refs}

\begin{thebibliography}{38}
\expandafter\ifx\csname natexlab\endcsname\relax\def\natexlab#1{#1}\fi
\expandafter\ifx\csname bibnamefont\endcsname\relax
  \def\bibnamefont#1{#1}\fi
\expandafter\ifx\csname bibfnamefont\endcsname\relax
  \def\bibfnamefont#1{#1}\fi
\expandafter\ifx\csname citenamefont\endcsname\relax
  \def\citenamefont#1{#1}\fi
\expandafter\ifx\csname url\endcsname\relax
  \def\url#1{\texttt{#1}}\fi
\expandafter\ifx\csname urlprefix\endcsname\relax\def\urlprefix{URL }\fi
\providecommand{\bibinfo}[2]{#2}
\providecommand{\eprint}[2][]{\url{#2}}

\bibitem[{\citenamefont{Novoselov et~al.}(2005)\citenamefont{Novoselov, Geim,
  Morozov, Jiang, Katsnelson, Grigorieva, Dubonos, and Firsov}}]{Novoselov}
\bibinfo{author}{\bibfnamefont{K.~S.} \bibnamefont{Novoselov}},
  \bibinfo{author}{\bibfnamefont{A.~K.} \bibnamefont{Geim}},
  \bibinfo{author}{\bibfnamefont{S.~V.} \bibnamefont{Morozov}},
  \bibinfo{author}{\bibfnamefont{D.}~\bibnamefont{Jiang}},
  \bibinfo{author}{\bibfnamefont{M.~I.} \bibnamefont{Katsnelson}},
  \bibinfo{author}{\bibfnamefont{I.~V.} \bibnamefont{Grigorieva}},
  \bibinfo{author}{\bibfnamefont{S.~V.} \bibnamefont{Dubonos}},
  \bibnamefont{and} \bibinfo{author}{\bibfnamefont{A.~A.}
  \bibnamefont{Firsov}}, \bibinfo{journal}{Nature (London)}
  \textbf{\bibinfo{volume}{438}}, \bibinfo{pages}{197} (\bibinfo{year}{2005}).

\bibitem[{\citenamefont{Geim and Novoselov}(2007)}]{Geim}
\bibinfo{author}{\bibfnamefont{A.~K.} \bibnamefont{Geim}} \bibnamefont{and}
  \bibinfo{author}{\bibfnamefont{K.~S.} \bibnamefont{Novoselov}},
  \bibinfo{journal}{Nat. Mater.} \textbf{\bibinfo{volume}{6}},
  \bibinfo{pages}{183} (\bibinfo{year}{2007}).

\bibitem[{\citenamefont{Yang et~al.}(2009)\citenamefont{Yang, Deslippe, Park,
  Cohen, and Louie}}]{Louie}
\bibinfo{author}{\bibfnamefont{L.}~\bibnamefont{Yang}},
  \bibinfo{author}{\bibfnamefont{J.}~\bibnamefont{Deslippe}},
  \bibinfo{author}{\bibfnamefont{C.-H.} \bibnamefont{Park}},
  \bibinfo{author}{\bibfnamefont{M.~L.} \bibnamefont{Cohen}}, \bibnamefont{and}
  \bibinfo{author}{\bibfnamefont{S.~G.} \bibnamefont{Louie}},
  \bibinfo{journal}{Phys. Rev. Lett.} \textbf{\bibinfo{volume}{103}},
  \bibinfo{pages}{186802} (\bibinfo{year}{2009}).

\bibitem[{\citenamefont{Maultzsch et~al.}(2004)\citenamefont{Maultzsch, Reich,
  Thomsen, Requardt, and Ordej{\'{o}}n}}]{Maultzsch}
\bibinfo{author}{\bibfnamefont{J.}~\bibnamefont{Maultzsch}},
  \bibinfo{author}{\bibfnamefont{S.}~\bibnamefont{Reich}},
  \bibinfo{author}{\bibfnamefont{C.}~\bibnamefont{Thomsen}},
  \bibinfo{author}{\bibfnamefont{H.}~\bibnamefont{Requardt}}, \bibnamefont{and}
  \bibinfo{author}{\bibfnamefont{P.}~\bibnamefont{Ordej{\'{o}}n}},
  \bibinfo{journal}{Phys. Rev. Lett.} \textbf{\bibinfo{volume}{92}},
  \bibinfo{pages}{075501} (\bibinfo{year}{2004}).

\bibitem[{\citenamefont{Cai et~al.}(2013)\citenamefont{Cai, Chuu, Wei, and
  Chou}}]{Chou}
\bibinfo{author}{\bibfnamefont{Y.}~\bibnamefont{Cai}},
  \bibinfo{author}{\bibfnamefont{C.-P.} \bibnamefont{Chuu}},
  \bibinfo{author}{\bibfnamefont{C.~M.} \bibnamefont{Wei}}, \bibnamefont{and}
  \bibinfo{author}{\bibfnamefont{M.~Y.} \bibnamefont{Chou}},
  \bibinfo{journal}{Phys. Rev. B} \textbf{\bibinfo{volume}{88}},
  \bibinfo{pages}{245408} (\bibinfo{year}{2013}).

\bibitem[{\citenamefont{P{\'{e}}rez et~al.}(2001)\citenamefont{P{\'{e}}rez,
  Ortega, and Flores}}]{Flores}
\bibinfo{author}{\bibfnamefont{R.}~\bibnamefont{P{\'{e}}rez}},
  \bibinfo{author}{\bibfnamefont{J.}~\bibnamefont{Ortega}}, \bibnamefont{and}
  \bibinfo{author}{\bibfnamefont{F.}~\bibnamefont{Flores}},
  \bibinfo{journal}{Phys. Rev. Lett.} \textbf{\bibinfo{volume}{86}},
  \bibinfo{pages}{4891} (\bibinfo{year}{2001}).

\bibitem[{\citenamefont{Marianetti and Yevick}(2010)}]{Yevick}
\bibinfo{author}{\bibfnamefont{C.~A.} \bibnamefont{Marianetti}}
  \bibnamefont{and} \bibinfo{author}{\bibfnamefont{H.~G.}
  \bibnamefont{Yevick}}, \bibinfo{journal}{Phys. Rev. Lett.}
  \textbf{\bibinfo{volume}{105}}, \bibinfo{pages}{245502}
  (\bibinfo{year}{2010}).

\bibitem[{\citenamefont{Si et~al.}(2012)\citenamefont{Si, Duan, Liu, and
  Liu}}]{FengLiu}
\bibinfo{author}{\bibfnamefont{C.}~\bibnamefont{Si}},
  \bibinfo{author}{\bibfnamefont{W.}~\bibnamefont{Duan}},
  \bibinfo{author}{\bibfnamefont{Z.}~\bibnamefont{Liu}}, \bibnamefont{and}
  \bibinfo{author}{\bibfnamefont{F.}~\bibnamefont{Liu}},
  \bibinfo{journal}{Phys. Rev. Lett.} \textbf{\bibinfo{volume}{109}},
  \bibinfo{pages}{226802} (\bibinfo{year}{2012}).

\bibitem[{\citenamefont{Krumhansl and Schrieffer}(1975)}]{Schrieffer}
\bibinfo{author}{\bibfnamefont{J.~A.} \bibnamefont{Krumhansl}}
  \bibnamefont{and} \bibinfo{author}{\bibfnamefont{J.~R.}
  \bibnamefont{Schrieffer}}, \bibinfo{journal}{Phys. Rev. B}
  \textbf{\bibinfo{volume}{11}}, \bibinfo{pages}{3535} (\bibinfo{year}{1975}).

\bibitem[{\citenamefont{Collins et~al.}(1979)\citenamefont{Collins, Blumen,
  Currie, and Ross}}]{Ross}
\bibinfo{author}{\bibfnamefont{M.~A.} \bibnamefont{Collins}},
  \bibinfo{author}{\bibfnamefont{A.}~\bibnamefont{Blumen}},
  \bibinfo{author}{\bibfnamefont{J.~F.} \bibnamefont{Currie}},
  \bibnamefont{and} \bibinfo{author}{\bibfnamefont{J.}~\bibnamefont{Ross}},
  \bibinfo{journal}{Phys. Rev. B} \textbf{\bibinfo{volume}{19}},
  \bibinfo{pages}{3630} (\bibinfo{year}{1979}).

\bibitem[{\citenamefont{B{\"{o}}ni et~al.}(1988)}]{Shirane}
\bibinfo{author}{\bibfnamefont{P.}~\bibnamefont{B{\"{o}}ni}}
  \bibnamefont{et~al.}, \bibinfo{journal}{Phys. Rev. B}
  \textbf{\bibinfo{volume}{38}}, \bibinfo{pages}{185} (\bibinfo{year}{1988}).

\bibitem[{\citenamefont{Ghosez et~al.}(1999)\citenamefont{Ghosez, Cockayne,
  Waghmare, and Rabe}}]{Ghosez}
\bibinfo{author}{\bibfnamefont{P.}~\bibnamefont{Ghosez}},
  \bibinfo{author}{\bibfnamefont{E.}~\bibnamefont{Cockayne}},
  \bibinfo{author}{\bibfnamefont{U.~V.} \bibnamefont{Waghmare}},
  \bibnamefont{and} \bibinfo{author}{\bibfnamefont{K.~M.} \bibnamefont{Rabe}},
  \bibinfo{journal}{Phys. Rev. B} \textbf{\bibinfo{volume}{60}},
  \bibinfo{pages}{836} (\bibinfo{year}{1999}).

\bibitem[{\citenamefont{Fawcett}(1988)}]{Cr}
\bibinfo{author}{\bibfnamefont{E.}~\bibnamefont{Fawcett}},
  \bibinfo{journal}{Rev. Mod. Phys.} \textbf{\bibinfo{volume}{60}},
  \bibinfo{pages}{209} (\bibinfo{year}{1988}).

\bibitem[{\citenamefont{Little}(1964)}]{Little}
\bibinfo{author}{\bibfnamefont{W.~A.} \bibnamefont{Little}},
  \bibinfo{journal}{Phys. Rev.} \textbf{\bibinfo{volume}{134}},
  \bibinfo{pages}{A1416} (\bibinfo{year}{1964}).

\bibitem[{\citenamefont{Kane and Mele}(2005)}]{Kane}
\bibinfo{author}{\bibfnamefont{C.~L.} \bibnamefont{Kane}} \bibnamefont{and}
  \bibinfo{author}{\bibfnamefont{E.~J.} \bibnamefont{Mele}},
  \bibinfo{journal}{Phys. Rev. Lett.} \textbf{\bibinfo{volume}{95}},
  \bibinfo{pages}{146802} (\bibinfo{year}{2005}).

\bibitem[{\citenamefont{Pesin and MacDonald}(2012)}]{MacDonald}
\bibinfo{author}{\bibfnamefont{D.}~\bibnamefont{Pesin}} \bibnamefont{and}
  \bibinfo{author}{\bibfnamefont{A.~H.} \bibnamefont{MacDonald}},
  \bibinfo{journal}{Nat. Mater.} \textbf{\bibinfo{volume}{11}},
  \bibinfo{pages}{409} (\bibinfo{year}{2012}).

\bibitem[{\citenamefont{Smeu et~al.}(2012)}]{Smeu}
\bibinfo{author}{\bibfnamefont{M.}~\bibnamefont{Smeu}} \bibnamefont{et~al.},
  \bibinfo{journal}{Phys. Rev. B} \textbf{\bibinfo{volume}{85}},
  \bibinfo{pages}{195315} (\bibinfo{year}{2012}).

\bibitem[{\citenamefont{Li et~al.}(2012)}]{Qikun}
\bibinfo{author}{\bibfnamefont{W.}~\bibnamefont{Li}} \bibnamefont{et~al.},
  \bibinfo{journal}{Nature Phys.} \textbf{\bibinfo{volume}{8}},
  \bibinfo{pages}{126} (\bibinfo{year}{2012}).

\bibitem[{\citenamefont{Chen et~al.}(2013)\citenamefont{Chen, Li, Feng, Ding,
  Qiu, Cheng, Wu, and Meng}}]{Chen}
\bibinfo{author}{\bibfnamefont{L.}~\bibnamefont{Chen}},
  \bibinfo{author}{\bibfnamefont{H.}~\bibnamefont{Li}},
  \bibinfo{author}{\bibfnamefont{B.}~\bibnamefont{Feng}},
  \bibinfo{author}{\bibfnamefont{Z.}~\bibnamefont{Ding}},
  \bibinfo{author}{\bibfnamefont{J.}~\bibnamefont{Qiu}},
  \bibinfo{author}{\bibfnamefont{P.}~\bibnamefont{Cheng}},
  \bibinfo{author}{\bibfnamefont{K.}~\bibnamefont{Wu}}, \bibnamefont{and}
  \bibinfo{author}{\bibfnamefont{S.}~\bibnamefont{Meng}},
  \bibinfo{journal}{Phys. Rev. Lett.} \textbf{\bibinfo{volume}{110}},
  \bibinfo{pages}{085504} (\bibinfo{year}{2013}).

\bibitem[{\citenamefont{Takeda and Shiraishi}(1994)}]{Takeda}
\bibinfo{author}{\bibfnamefont{K.}~\bibnamefont{Takeda}} \bibnamefont{and}
  \bibinfo{author}{\bibfnamefont{K.}~\bibnamefont{Shiraishi}},
  \bibinfo{journal}{Phys. Rev. B} \textbf{\bibinfo{volume}{50}},
  \bibinfo{pages}{14916} (\bibinfo{year}{1994}).

\bibitem[{\citenamefont{Cahangirov et~al.}(2009)\citenamefont{Cahangirov,
  Topsakal, Akt{\"{u}}rk, {\c{S}}ahin, and Ciraci}}]{Ciraci}
\bibinfo{author}{\bibfnamefont{S.}~\bibnamefont{Cahangirov}},
  \bibinfo{author}{\bibfnamefont{M.}~\bibnamefont{Topsakal}},
  \bibinfo{author}{\bibfnamefont{E.}~\bibnamefont{Akt{\"{u}}rk}},
  \bibinfo{author}{\bibfnamefont{H.}~\bibnamefont{{\c{S}}ahin}},
  \bibnamefont{and} \bibinfo{author}{\bibfnamefont{S.}~\bibnamefont{Ciraci}},
  \bibinfo{journal}{Phys. Rev. Lett.} \textbf{\bibinfo{volume}{102}},
  \bibinfo{pages}{236804} (\bibinfo{year}{2009}).

\bibitem[{\citenamefont{Lin et~al.}(2012)}]{Kawai}
\bibinfo{author}{\bibfnamefont{C.-L.} \bibnamefont{Lin}} \bibnamefont{et~al.},
  \bibinfo{journal}{Appl. Phys. Express} \textbf{\bibinfo{volume}{5}},
  \bibinfo{pages}{045802} (\bibinfo{year}{2012}).

\bibitem[{\citenamefont{Vogt et~al.}(2012)}]{Vogt}
\bibinfo{author}{\bibfnamefont{P.}~\bibnamefont{Vogt}} \bibnamefont{et~al.},
  \bibinfo{journal}{Phys. Rev. Lett.} \textbf{\bibinfo{volume}{108}},
  \bibinfo{pages}{155501} (\bibinfo{year}{2012}).

\bibitem[{\citenamefont{Fleurence et~al.}(2012)}]{Fleurence}
\bibinfo{author}{\bibfnamefont{A.}~\bibnamefont{Fleurence}}
  \bibnamefont{et~al.}, \bibinfo{journal}{Phys. Rev. Lett.}
  \textbf{\bibinfo{volume}{108}}, \bibinfo{pages}{245501}
  (\bibinfo{year}{2012}).

\bibitem[{\citenamefont{Feng et~al.}(2012)}]{Feng}
\bibinfo{author}{\bibfnamefont{B.}~\bibnamefont{Feng}} \bibnamefont{et~al.},
  \bibinfo{journal}{Nano Lett.} \textbf{\bibinfo{volume}{12}},
  \bibinfo{pages}{3507} (\bibinfo{year}{2012}).

\bibitem[{\citenamefont{Meng et~al.}(2013)}]{Meng}
\bibinfo{author}{\bibfnamefont{L.}~\bibnamefont{Meng}} \bibnamefont{et~al.},
  \bibinfo{journal}{Nano Lett.} \textbf{\bibinfo{volume}{13}},
  \bibinfo{pages}{685} (\bibinfo{year}{2013}).

\bibitem[{\citenamefont{Lee et~al.}(2013)\citenamefont{Lee, Fleurence,
  Friedlein, Yamada-Takamura, and Ozaki}}]{Chicheng}
\bibinfo{author}{\bibfnamefont{C.-C.} \bibnamefont{Lee}},
  \bibinfo{author}{\bibfnamefont{A.}~\bibnamefont{Fleurence}},
  \bibinfo{author}{\bibfnamefont{R.}~\bibnamefont{Friedlein}},
  \bibinfo{author}{\bibfnamefont{Y.}~\bibnamefont{Yamada-Takamura}},
  \bibnamefont{and} \bibinfo{author}{\bibfnamefont{T.}~\bibnamefont{Ozaki}},
  \bibinfo{journal}{Phys. Rev. B} \textbf{\bibinfo{volume}{88}},
  \bibinfo{pages}{165404} (\bibinfo{year}{2013}).

\bibitem[{\citenamefont{Friedlein et~al.}(2013)\citenamefont{Friedlein,
  Fleurence, Sadowski, and Yamada-Takamura}}]{Friedlein}
\bibinfo{author}{\bibfnamefont{R.}~\bibnamefont{Friedlein}},
  \bibinfo{author}{\bibfnamefont{A.}~\bibnamefont{Fleurence}},
  \bibinfo{author}{\bibfnamefont{J.~T.} \bibnamefont{Sadowski}},
  \bibnamefont{and}
  \bibinfo{author}{\bibfnamefont{Y.}~\bibnamefont{Yamada-Takamura}},
  \bibinfo{journal}{Appl. Phys. Lett.} \textbf{\bibinfo{volume}{102}},
  \bibinfo{pages}{221603} (\bibinfo{year}{2013}).

\bibitem[{\citenamefont{Lee et~al.}(2014)\citenamefont{Lee, Fleurence,
  Yamada-Takamura, Ozaki, and Friedlein}}]{Chicheng2}
\bibinfo{author}{\bibfnamefont{C.-C.} \bibnamefont{Lee}},
  \bibinfo{author}{\bibfnamefont{A.}~\bibnamefont{Fleurence}},
  \bibinfo{author}{\bibfnamefont{Y.}~\bibnamefont{Yamada-Takamura}},
  \bibinfo{author}{\bibfnamefont{T.}~\bibnamefont{Ozaki}}, \bibnamefont{and}
  \bibinfo{author}{\bibfnamefont{R.}~\bibnamefont{Friedlein}}
  (\bibinfo{year}{2014}), \bibinfo{note}{http://arxiv.org/abs/1407.2698}.

\bibitem[{\citenamefont{Wang et~al.}(2013)\citenamefont{Wang, Zhang,
  Liz{\'{a}}rraga, Marco, and Eriksson}}]{Eriksson}
\bibinfo{author}{\bibfnamefont{B.-T.} \bibnamefont{Wang}},
  \bibinfo{author}{\bibfnamefont{P.}~\bibnamefont{Zhang}},
  \bibinfo{author}{\bibfnamefont{R.}~\bibnamefont{Liz{\'{a}}rraga}},
  \bibinfo{author}{\bibfnamefont{I.~D.} \bibnamefont{Marco}}, \bibnamefont{and}
  \bibinfo{author}{\bibfnamefont{O.}~\bibnamefont{Eriksson}},
  \bibinfo{journal}{Phys. Rev. B} \textbf{\bibinfo{volume}{88}},
  \bibinfo{pages}{104107} (\bibinfo{year}{2013}).

\bibitem[{\citenamefont{Morrison et~al.}(1993)\citenamefont{Morrison, Bylander,
  and Kleinman}}]{MBK}
\bibinfo{author}{\bibfnamefont{I.}~\bibnamefont{Morrison}},
  \bibinfo{author}{\bibfnamefont{D.}~\bibnamefont{Bylander}}, \bibnamefont{and}
  \bibinfo{author}{\bibfnamefont{L.}~\bibnamefont{Kleinman}},
  \bibinfo{journal}{Phys. Rev. B} \textbf{\bibinfo{volume}{47}},
  \bibinfo{pages}{6728} (\bibinfo{year}{1993}).

\bibitem[{\citenamefont{Ozaki}(2003)}]{Ozaki}
\bibinfo{author}{\bibfnamefont{T.}~\bibnamefont{Ozaki}},
  \bibinfo{journal}{Phys. Rev. B} \textbf{\bibinfo{volume}{67}},
  \bibinfo{pages}{155108} (\bibinfo{year}{2003}).

\bibitem[{\citenamefont{Ozaki et~al.}()}]{openmx}
\bibinfo{author}{\bibfnamefont{T.}~\bibnamefont{Ozaki}} \bibnamefont{et~al.},
  \urlprefix\url{http://www.openmx-square.org/}.

\bibitem[{\citenamefont{Ohwaki et~al.}(2012)\citenamefont{Ohwaki, Otani,
  Ikeshoji, and Ozaki}}]{Ohwaki}
\bibinfo{author}{\bibfnamefont{T.}~\bibnamefont{Ohwaki}},
  \bibinfo{author}{\bibfnamefont{M.}~\bibnamefont{Otani}},
  \bibinfo{author}{\bibfnamefont{T.}~\bibnamefont{Ikeshoji}}, \bibnamefont{and}
  \bibinfo{author}{\bibfnamefont{T.}~\bibnamefont{Ozaki}}, \bibinfo{journal}{J.
  Chem. Phys.} \textbf{\bibinfo{volume}{136}}, \bibinfo{pages}{134101}
  (\bibinfo{year}{2012}).

\bibitem[{\citenamefont{Kohn and Sham}(1965)}]{Kohn}
\bibinfo{author}{\bibfnamefont{W.}~\bibnamefont{Kohn}} \bibnamefont{and}
  \bibinfo{author}{\bibfnamefont{L.~J.} \bibnamefont{Sham}},
  \bibinfo{journal}{Phys. Rev.} \textbf{\bibinfo{volume}{140}},
  \bibinfo{pages}{A1133} (\bibinfo{year}{1965}).

\bibitem[{\citenamefont{Perdew et~al.}(1996)\citenamefont{Perdew, Burke, and
  Ernzerhof}}]{Perdew}
\bibinfo{author}{\bibfnamefont{J.~P.} \bibnamefont{Perdew}},
  \bibinfo{author}{\bibfnamefont{K.}~\bibnamefont{Burke}}, \bibnamefont{and}
  \bibinfo{author}{\bibfnamefont{M.}~\bibnamefont{Ernzerhof}},
  \bibinfo{journal}{Phys. Rev. Lett.} \textbf{\bibinfo{volume}{77}},
  \bibinfo{pages}{3865} (\bibinfo{year}{1996}).

\bibitem[{\citenamefont{Kokalj}(2003)}]{xcrysden}
\bibinfo{author}{\bibfnamefont{A.}~\bibnamefont{Kokalj}},
  \bibinfo{journal}{Comp. Mater. Sci.} \textbf{\bibinfo{volume}{28}},
  \bibinfo{pages}{155} (\bibinfo{year}{2003}).

\bibitem[{\citenamefont{Yamada-Takamura
  et~al.}(2010)\citenamefont{Yamada-Takamura, Bussolotti, Fleurence, Bera, and
  Friedlein}}]{Takamura}
\bibinfo{author}{\bibfnamefont{Y.}~\bibnamefont{Yamada-Takamura}},
  \bibinfo{author}{\bibfnamefont{F.}~\bibnamefont{Bussolotti}},
  \bibinfo{author}{\bibfnamefont{A.}~\bibnamefont{Fleurence}},
  \bibinfo{author}{\bibfnamefont{S.}~\bibnamefont{Bera}}, \bibnamefont{and}
  \bibinfo{author}{\bibfnamefont{R.}~\bibnamefont{Friedlein}},
  \bibinfo{journal}{Appl. Phys. Lett.} \textbf{\bibinfo{volume}{97}},
  \bibinfo{pages}{073109} (\bibinfo{year}{2010}).

\end{thebibliography}
\end{document}